\documentclass[preprint, showpacs,preprintnumbers,amsmath,amssymb,nofootinbib]{revtex4}
\usepackage{amssymb}

\usepackage{graphicx}
\usepackage{dcolumn}
\usepackage{bm}

\newcommand{\bea}{\begin{eqnarray}}
\newcommand{\eea}{\end{eqnarray}}

\begin{document}
\title{ Probing  the cosmic acceleration history and the properties of dark energy
from the ESSENCE supernova data with a model independent method}
\author{  Puxun Wu\;$^{1,2}$ and  Hongwei Yu\;$^{3}$
}

\address
{$^1$Department of Physics and Tsinghua Center for Astrophysics,
Tsinghua University, Beijing 100084, China
\\
$^2$Institute of  Math-Physics and School  of Sciences, Central
South University of Forestry and Technology, Changsha, Hunan 410004,
China
 \\
$^3$Department of Physics and Institute of  Physics,\\ Hunan Normal
University, Changsha, Hunan 410081, China
}

\begin{abstract}
With a model independent method the expansion history $H(z)$, the
deceleration parameter $q(z)$  of the universe and the equation of
state $w(z)$ for the dark energy are reconstructed directly from the
192 Sne Ia data points, which contain the new ESSENCE Sne Ia data
and the high redshift Sne Ia data. We find that the evolving
properties of $q(z)$ and $w(z)$ reconstructed from the 192 Sne Ia
data seem to be weaker than that obtained from the Gold set, but
stronger than that from the SNLS set. With a combination of the 192
Sne Ia and BAO data, a tight constraint on $\Omega_{m0}$ is
obtained. At the $1\sigma$ confidence level
$\Omega_{m0}=0.278^{+0.024}_{-0.023}$, which is highly consistent
with that from the Gold+BAO and SNLS+BAO.
\end{abstract}


 \maketitle

\section{Introduction}
In order to explain the present cosmic accelerating expansion
discovered firstly from the Type Ia Supernovae  (Sne
Ia)~\cite{Perlmutter1999, Riess1998, Riess2004, Riess2007,
Astier2006, Wood2007},  dark energy (see \cite{Padmanabhan2006,
Copeland2006, Sahni2006, Perivolaropoulos2006}  for recent reviews)
is usually assumed to exist in the universe. Dark energy is  an
exotic energy component with negative pressure, and presumably began
to dominate the evolution of the universe only recently. Although it
has been studied for nearly one decade, its nature is still
puzzling.

In general there are two kinds of methods to reconstruct the
properties of dark energy from the observation data directly. One is
to assume an arbitrary parametrization for the equation of state of
dark energy, $w(z)$,   the potential of dark energy, $V(z)$, the
Hubble parameter, $H(z)$, or the luminosity distance $d_L(z) $ with
some arbitrary constants. By determining these constants from the
observational data, we can obtain the evolving properties of dark
energy. Different ways of parametrization have been discussed in
 Refs.~\cite{line1,Sahni2003, Wetterich2004}. The other is the
no-parametric method, which usually involves directly smoothing
either $d_L$, or some other quantity  with some characteristic
smoothing scale. Currently there are many different models of
implementing this approach~\cite{Wang2001, Shaf2006, Shaf2007}

Recently, based on smoothing the noise of supernova data over
redshift, the authors in Refs.~\cite{Shaf2006, Shaf2007} suggested a
no-parametric method in a model independent manner to reconstruct
the expansion history of our universe and the evolving properties of
dark energy. In Ref.~\cite{Shaf2007} two kinds of supernova data:
182 Gold dataset and 115 SNLS dataset are used firstly to
reconstruct the Hubble parameter $h(z)$ ($h(z)=H(z)/H_0$) and the
deceleration parameter $q(z)$. It was found that  both data sets
give $q(0)<0$,  which means the universe is undergoing an
accelerating expansion, while the Gold set seems to favor a later
entering of this accelerating era than the SNLS one. However Gold
and SNLS give a good consistent constraint on the present matter
density parameter $\Omega_{m0}$ ($\Omega_{m0}\approx 0.276\pm
0.023$) when combined with the baryonic acoustic oscillation peak
obtained from the large scale correlation function of luminous red
galaxy in the Sloan Digital Sky Survey (SDSS) \cite{Eisenstein2005}.
The $w(z)$ was also discussed with both kinds of SNe Ia datasets,
and it was found, in agreement with that obtained in Refs.~
\cite{Nesseris2006,Alam2006, WuYu2007} using some parameterized
models, the Gold slightly favors an dynamically evolving dark energy
with a crossing of phantom divide line while the SNLS does not.
However, in this method, the present value of Hubble parameter,
$H_0$, is needed  prior or should be marginalized over. Since the
value of $H_0$ from different observation data seems to be
inconsistent and  doing the marginalization wastes the compute
resources,  in this paper, we firstly generalize this model
independent approach to eliminate the impact of $H_0$ and then
reconstruct the cosmic expansion history, $H(z)$, the deceleration
parameter, $q(z)$, and the equation of state for dark energy $w(z)$
from the new ESSENCE Sne Ia data. Beside the 162 data points given
in table 9 in Ref. \cite{Wood2007}, which contains 60 ESSENCE Sne
Ia, 57 SNLS Sne Ia and  45 nearby Sne Ia, we add 30 Sne Ia detected
at $0.216<z<1.755$ by the Hubble Space Telescope~\cite{Riess2007}.

\section{The method}
Following a well known procedure in the analysis of large scale
structure,  Shafieloo et al. \cite{Shaf2006, Shaf2007} use a
Gaussian smoothing function rather than the top hat smoothing
function to smooth the noise of the Sne Ia data directly. In order
to obtain the important information of interested cosmological
parameters expediently,  $\ln d_L(z)$ rather than the luminosity
distance $d_L(z)$ or distance module $\mu(z)$ is studied by the
following iterative method
\begin{eqnarray}
\ln d_L(z)^s_n=\ln d_L(z)^s_{n-1}+N(z)\sum_i(\ln d^{obs}_L(z_i)-\ln
d_L(z_i)^s_{n-1})\exp\bigg[-{\ln^2\big({1+z \over 1+z_i}\big)\over
2\triangle^2}\bigg]\;,
\end{eqnarray}
with a normalization parameter
\begin{eqnarray}
N(z)^{-1}=\sum_i \exp\bigg[-{\ln^2\big({1+z \over 1+z_i}\big)\over
2\triangle^2}\bigg]\;.
\end{eqnarray}
In Eqs.(1,2)  $\triangle$ is a quantity needed to be given prior.
Since a large value of $\triangle$ leads to a smooth result but
depresses the accuracy of reconstruction, and inversely for a small
value of  $\triangle$. So it is important to choose a reasonable
value of $\triangle$. Here, as in Ref. \cite{Shaf2007}, we choose
$\triangle=0.6$. In Eq.(1),  $d_L(z)^s_n$ represents the smoothed
luminosity distance at any redshift $z$ after $n$ iteration. When
$n=1$  $d_L(z)^s_{0}$ denotes a guess background model and it has
been shown that the results are not sensitive to the chosen vaule of
$\Delta$ and the assumed initial guess model~\cite{Shaf2007}. In
this paper we use a $wCDM$ model with $w=-0.9$ and
$\Omega_{m0}=0.28$ as this guess background model.  $\ln d^{obs}_L
(z_i)$ is the observed one from the Sne Ia and can be expressed as:
\begin{eqnarray}
\ln d^{obs}_L (z_i)={\ln 10\over 5} [\mu^{obs}(z_i)-42.38]+\ln
h\equiv \ln f^{obs}(z_i)+\ln h\;.
\end{eqnarray}
Here $h=H_0/100$ and $\mu^{obs}$ is the observed distance module of
Sne Ia. Apparently using the above method the nuisance parameter $h$
needs to be given prior or marginalized over. Now we generalize this
method to eliminate the impact of $h$.  Substituting Eq. (3) into
Eq. (1), we obtain that
\begin{eqnarray}
\ln d_L(z)^s_n=\ln d_L(z)^s_{n-1}+N(z)\sum_i(\ln f^{obs}(z_i)-\ln
d_L(z_i)^s_{n-1})\exp\bigg[-{\ln^2\big({1+z \over 1+z_i}\big)\over
2\triangle^2}\bigg]+\ln h\;.
\end{eqnarray}
If defining $\ln d_L(z)^s_n=\ln f(z)^s_n+\ln h$, it is easy to see
\begin{eqnarray}
\ln f(z)^s_n=\ln f(z)^s_{n-1}+N(z)\sum_i(\ln f^{obs}(z_i)-\ln
f(z)^s_{n-1})\exp\bigg[-{\ln^2\big({1+z \over 1+z_i}\big)\over
2\triangle^2}\bigg]\;.
\end{eqnarray}
When $n=1$
\begin{eqnarray}
\ln f(z)^s_1 &=& \ln f(z)^s_{0}+N(z)\sum_i(\ln f^{obs}(z_i)-\ln
f(z)^s_{0})\exp\bigg[-{\ln^2\big({1+z \over 1+z_i}\big)\over
2\triangle^2}\bigg]\nonumber\\  &=&\ln d_L(z)^s_{0}+N(z)\sum_i(\ln
f^{obs}(z_i)-\ln d_L(z)^s_{0})\exp\bigg[-{\ln^2\big({1+z \over
1+z_i}\big)\over 2\triangle^2}\bigg]\;.
\end{eqnarray}
Here $d_L(z)^s_{0}$ is the luminosity distance of the suggested
background model. Different from Ref.\cite{Shaf2007} to iterate
through Eq.(1),   we use  Eq. (5) to obtain the smoothed results.
The advantage of doing so is that result is  independent of $h$ (or
$H_0$). In order to determine whether we obtain a best fit model
after some iteration, we calculate, after each iteration, $\chi^2$:
\begin{eqnarray}
\chi^2_n=\sum_i {(\mu(z_i)_n-\mu^{obs}(z_i))^2 \over
\sigma^2_{\mu_{obs,i}}}\;.
\end{eqnarray}
Once this $\chi^2_n$ reaches its minimum value we stop the iterative
process and get the best fit result.

By differentiating the smoothed luminosity distance we can find the
Hubble parameter,  $H(z)$, (not $h(z)$)
\begin{eqnarray}
H(z)=\bigg[{d\over dz}\bigg({100 f(z) \over 1+z}\bigg)\bigg]^{-1}\;,
\end{eqnarray}
which contains the information of $H_0$. Then the deceleration
parameter $q(z)$ of the universe and the equation of state $w(z)$ of
dark energy can be obtained:
\begin{eqnarray}
q(z)=(1+z){H'(z) \over H(z)}-1\;,
\end{eqnarray}
\begin{eqnarray}
w(z)={[2(1+z)/3]H'/H-1 \over 1-(H_0/H)^2\Omega_{m0}(1+z)^2}\;.
\end{eqnarray}

\section{The results}
The ESSENCE program (Equation of State: Supernovae trace Cosmic
Expansion¡ªan NOAO Survey Program) is designed to measure the
history of cosmic expansion over the past 5 billion years. The four
year data was released in Ref. \cite{Wood2007}, which contains $60$
Sne Ia points. Here we use the 162 data points  given in table 9 in
Ref. \cite{Wood2007}, which contains 60 ESSENCE Sne Ia, 57 SNLS Sne
Ia and  45 nearby Sne Ia. In addition, as in Ref.~\cite{Davis2007},
we add 30 Sne Ia detected at $0.216<z<1.755$ by the Hubble Space
Telescope~\cite{Riess2007}.

Using these 192 Sne Ia data points, we find when $n=42$ a minimum
value of $\chi^2$ is obtained which can be seen from the Fig. (1).
In Fig. (2) we show the reconstructed result of the Hubble parameter
$H(z)$ with the likelihood within $1 \sigma$. The red line is the
best fit result and when $z=0$ the best fit value of $H_0$ is
$H_0=65.5$. Fig. (3) shows the evolving curves of reconstructed
$q(z)$ with $1\sigma$ error bar. It is easy to see that  the
universe is undergoing an accelerating expansion since the present
value of $q(z)$ is less than zero, and the phase transition from
deceleration to acceleration occurs  at redshift $z\sim 0.55-0.73$
within $1 \sigma$, which is  slightly later than that obtained from
Gold set ($z\sim 0.38-0.48$) but earlier than that from SNLS set
($z>0.7$) \cite{Shaf2007}.

Fig. (4) shows the constraint on the present matter density
parameter $\Omega_{m0}$ with $H_0=65.5$ by combining  the ESSENCE
Sne Ia and baryonic acoustic oscillation(BAO) peak obtained from the
large scale correlation function of luminous red galaxy in the Sloan
Digital Sky Survey (SDSS). For BAO data we use  a model-independent
dimensionless parameter $A$ defined as
\begin{eqnarray}
A=\frac{\sqrt{\Omega_{m0}}}{h(z_1)^{1/3}}\bigg[\frac{1}{z_1}
  \int_0^{z_1}\frac{dz}{h(z)}\bigg]^{2/3}\;,
\end{eqnarray}
for a flat universe,  where $z_1 = 0.35$ and $A$ is measured to be
$A = 0.469({n\over 0.96})^{-0.35} \pm 0.017$ \cite{Eisenstein2005}.
Here $n$ is the spectral index of the primordial power spectrum and
the WMAP3 gives $n=0.951$ \cite{WMAP3}. Clearly the ESSENCE Sne Ia
and BAO give a strong constraint on $\Omega_{m0}$. At the $1\sigma$
confidence level we obtain $\Omega_{m0}=0.278^{+0.024}_{-0.023}$,
which is highly consistent with that obtained from Gold+BAO and
SNLS+BAO ( $\Omega_{m0}=0.276\pm 0.023$)~\cite{Shaf2007}.

In Fig. (5) we plot the evolving behavior of $w(z)$ with a
marginalization of $\Omega_{m0}$ over
$\Omega_{m0}=0.278^{+0.024}_{-0.023}$. The best fit (red) line shows
that the ESSENCE data slightly favors an evolving dark energy with a
crossing of phantom divide line at the near past, however this
evolving property is weaker than that obtained from Gold data but
stronger than that from SNSL set obtained in Ref.~\cite{Shaf2007}.
In addition, from  Figs. (3) and (5) we find that  the stringent
constraint on $w(z)$ and $q(z)$ happens around redshift $z\sim 0.5$,
which is consistent with that obtained in Refs.~\cite{Gong,
WuYu2007} with some parameterized models.

\section{conclusion}
In this paper, with a model independent method  we have
reconstructed the cosmic expansion history and the properties of
dark energy from recent ESSENCE Sne Ia data. We firstly obtain the
evolution of $H(z)$. Then the cosmic deceleration parameter $q(z)$
and the equation of state $w(z)$ of dark energy are reconstructed.
Our results show that their evolutionary property reconstructed from
ESSENCE data is weaker than that from Gold set, but is stronger than
that from the SNSL one. Combining the ESSENCE Sne Ia and the BAO
data, a tight constraint on $\Omega_{m0}$ is obtained. At a
$1\sigma$ confidence level $\Omega_{m0}=0.278^{+0.024}_{-0.023}$,
which is highly consistent with that obtained from Gold+BAO and
SNLS+BAO. Remarkably, as that obtained with some parameterized
model~\cite{ Gong, WuYu2007}, the tight constraints on $w(z)$ and
$q(z)$ seem to happen at about $z\sim 0.5$.

\begin{acknowledgments}
This work was supported in part by the National Natural Science
Foundation of China under Grants No. 10575035, 10775050 and
10705055, the Program for NCET under Grant No. 04-0784, the SRFDP
under Grant No. 20070542002, the Youth Scientific Research Fund of
Hunan Provincial Education Department under Grant No. 07B085, and
the Foundation of CSUFT under Grant No. 06Y020.
\end{acknowledgments}

\begin{figure}[htbp]
\includegraphics[width=9cm]{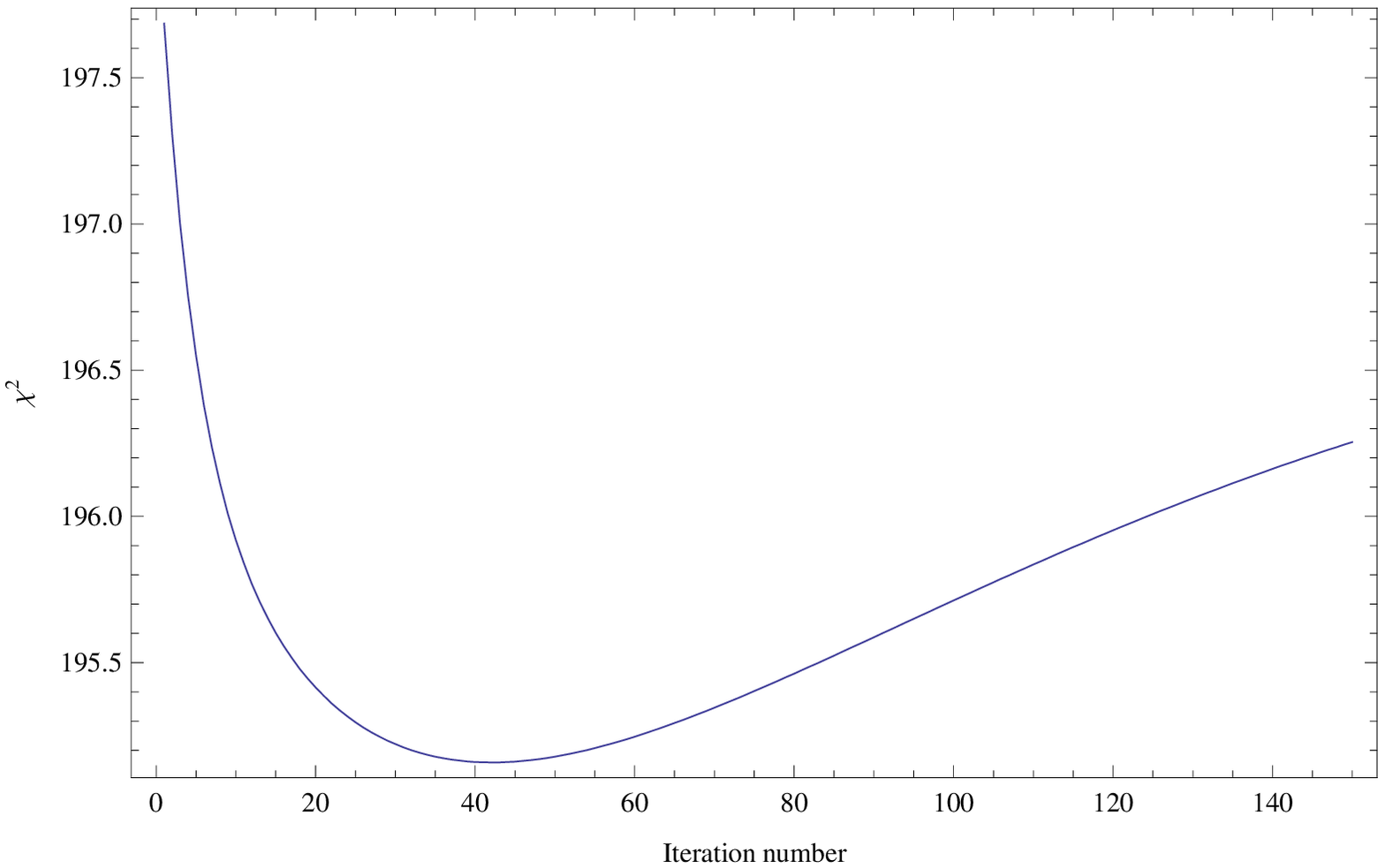}
\caption{\label{Fig1} Computed $\chi^2$ for the reconstructed
results at each iteration for the 192 Sne Ia data.}
\end{figure}

\begin{figure}[htbp]
\includegraphics[width=9cm]{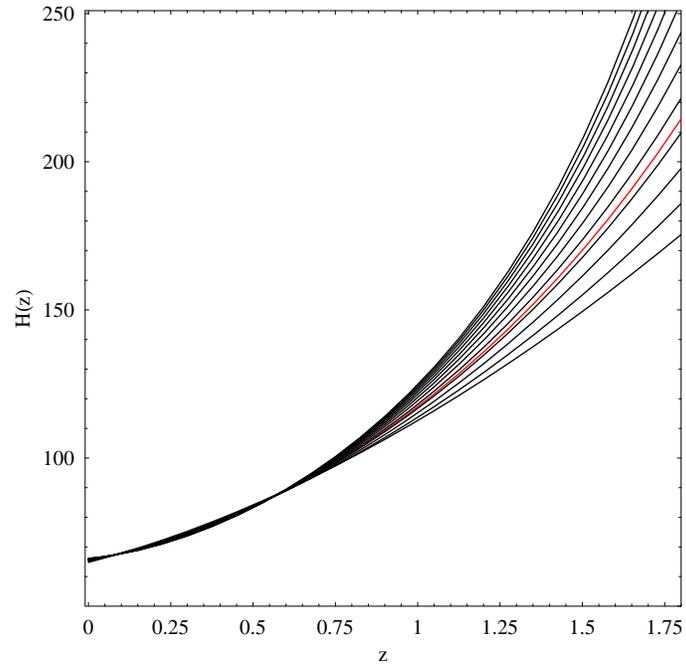}
\caption{\label{Fig2} The reconstructed evolutionary curves of the
Hubble parameter $H(z)$ with the likelihood within $1\sigma$. The
red line is the best recovered result.
}
\end{figure}

\begin{figure}[htbp]
\includegraphics[width=9cm]{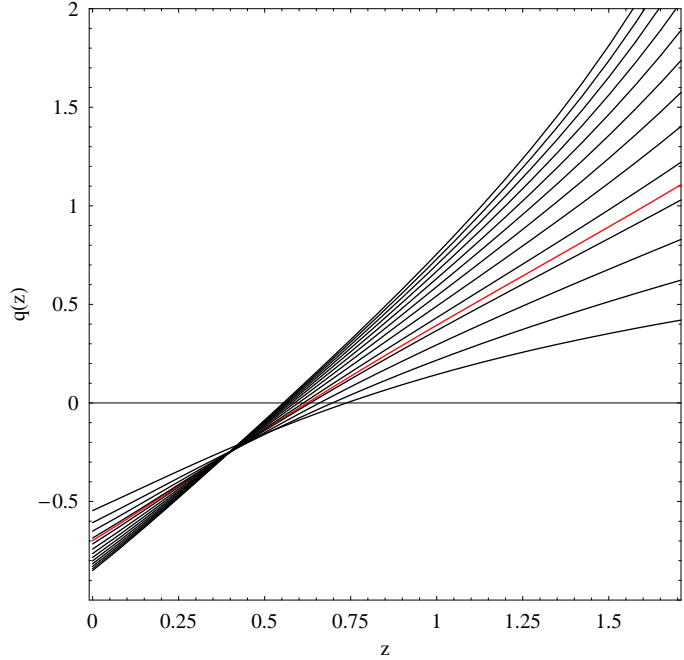}
\caption{\label{Fig3} The reconstructed evolutionary curves of the
deceleration parameter $q(z)$ with the likelihood within $1\sigma$.
The red line is the best recovered result. }
\end{figure}

\begin{figure}[htbp]
\includegraphics[width=9cm]{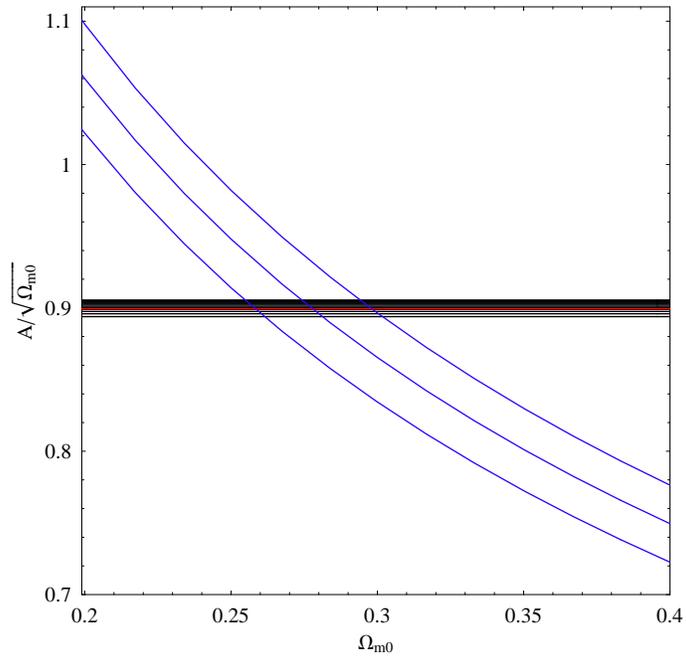}
\caption{\label{Fig4}The constraint on $\Omega_{m0}$ from the
combination of 192 Sne Ia and BAO data. The red and black lines show
the derived value of $A/\sqrt{\Omega_{m0}}$ from the 192 Sne Ia
dataset within $1\sigma$ and the blue lines are the results from the
BAO data. }
\end{figure}

\begin{figure}[htbp]
\includegraphics[width=9cm]{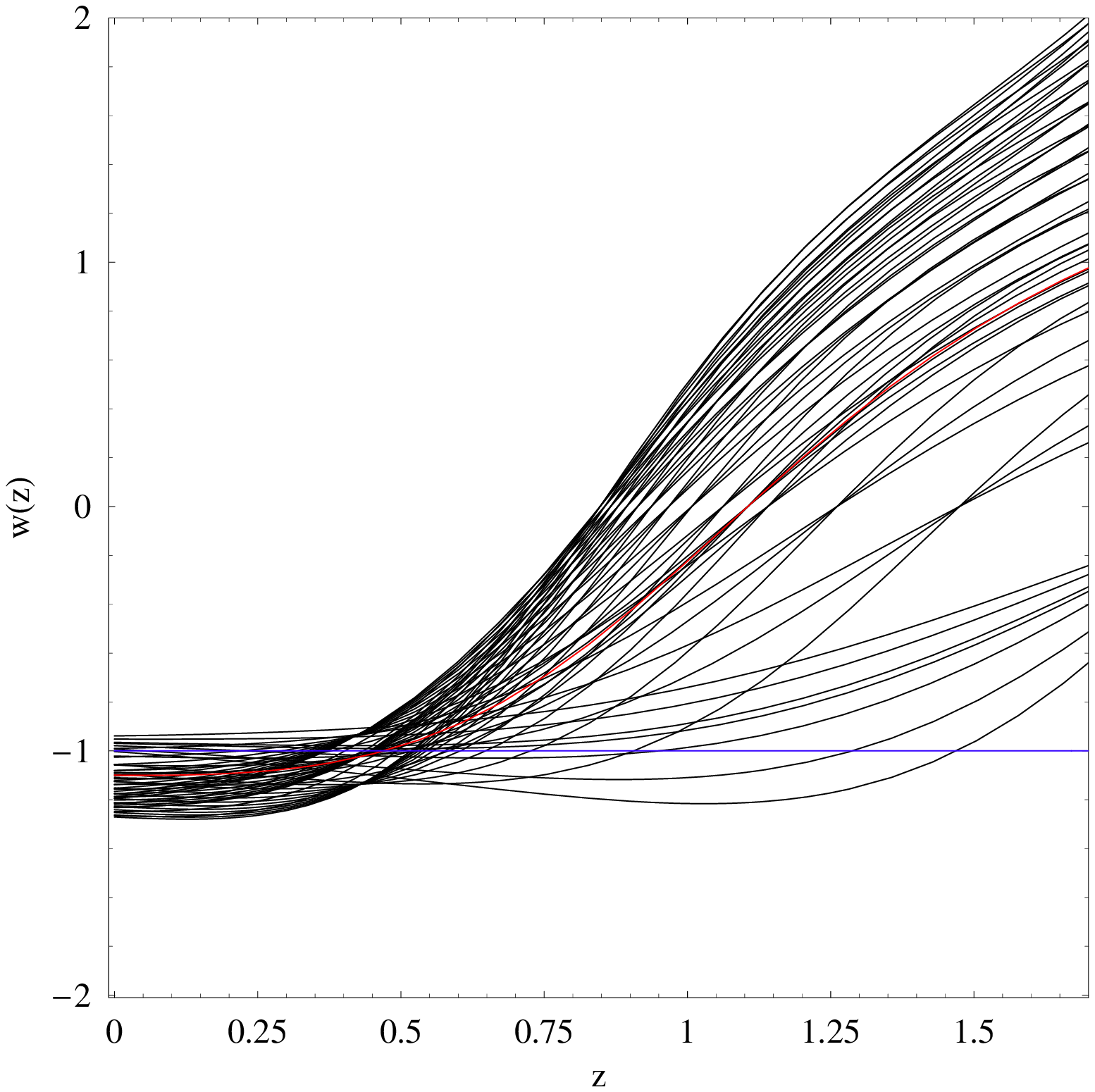}
\caption{\label{Fig5}The reconstructed evolutionary curves of the
equation of state of dark energy, $w(z)$, within $1\sigma$ with a
marginalization over $\Omega_{m0}=0.278^{+0.024}_{-0.023}$.}
\end{figure}

\begin{thebibliography}{99}
\bibitem{Perlmutter1999}S. Perlmutter, et al., 1999 Astrophys. J. {\bf 517} 565
\bibitem{Riess1998}     A. G. Riess,   et al., 1998 Astron. J. {\bf 116} 1009
\bibitem{Riess2004}     A. G. Riess,  et al.,  2004 Astrophys. J. {\bf 607} 665
\bibitem{Riess2007}     A. G. Riess,  et al., 2007 Astrophys. J. {\bf 659} 98
\bibitem{Astier2006}    P. Astier et al., 2006  Astron. Astrophys. {\bf 447} 31  
\bibitem{Wood2007}      W. M. Wood-Vasey et al., 2007 arXiv: astro-ph/0701041
\bibitem{Padmanabhan2006}T. Padmanabhan, 2006  AIP Conf. Proc. {\bf 861} 179 
\bibitem{Copeland2006}  E. J. Copeland, M. Sami  and S. Tsujikawa, 2006 Int. J. Mod. Phys. D {\bf 15} 1753
\bibitem{Sahni2006}     V. Sahni  and A. Starobinsky, 2006 Int. J. Mod. Phys. D {\bf 15} 2105 
\bibitem{Perivolaropoulos2006}L. Perivolaropoulos, 2006 arXiv: astro-ph/0601014
\bibitem{line1} A. A. Starobinsky, 1998 JETP Lett. {\bf 68} 757\\
               D. Huterer and M. S. Turner, 1999 Phys. Rev. D {\bf 60} 081301\\
               P. Astier, 2000 arXiv:astro-ph/0008306\\
               T. Chiba and T. Nakamura, 2000 Phys. Rev. D {\bf 62} 121301\\
               J. Weller and A. Albrecht, 2002 Phys. Rev. D {\bf 65} 103512 \\
               I. Maor, R. Brustein, J. McMahon and P. J. Steinhardt, 2002 Phys. Rev. D {\bf 65} 123003\\
               M. Chevallier, D. Polarski, 2001 Int. J. Mod. Phy. D. {\bf  10} 213 \\
               E. V. Linder, 2003 Phys. Rev. Lett. {\bf 90} 091301\\
               H. K. Jassal, J .S. Bagla and T. Padmanabhan, 2005 Mon. Not. Roy. Astron. Soc. {\bf 356} L11
\bibitem{Sahni2003} V. Sahni, T. D. Saini, A. A. Starobinsky and U. Alam, 2003 JETP lett. {\bf 77} 201\\ 
                    U. Alam,  V. Sahni, T. D. Saini and A. A. Starobinsky, 2003 Mon. not. Roy. Ast. Soc. {\bf 344} 1057 \\ 
                    U. Alam, V. Sahni, T. D. Saini and A. A. Starobinsky, 2004 astro-ph/0406672
                    T. D. Saini, S. Raychaudhury, V. Sahni and A. A. Starobinsky, 2000 Phys. Rev. Lett. {\bf 85} 1162 \\
                    T. D. Saini, J. Weller and S. L. Bridle, 2004 Mon. Not. Roy. Ast. Soc. {\bf 348} 603\\
                    U. Alam, V. Sahni and A. A. Starobinsky, 2004 J. Cosmol. Astropart.Phys. {\bf 0406}  008\\
                    U. Alam, V. Sahni, T. D. Saini and A. A. Starobinsky,  2004 Mon. Not. Roy. Ast. Soc. {\bf 354} 275 
\bibitem{Wetterich2004}B. Gerke and G. Efstathiou, 2002 Mon. Not. Roy. Ast. Soc. {\bf 335} 33 \\
                       I. Maor, R. Brustein, J. McMahon and P. J. Steinhardt, 2002 Phys. Rev. D {\bf 65} 123003 \\
                       P. S. Corasaniti and E. J. Copeland, 2003 Phys. Rev. D {\bf 67} 063521 \\
                       Y. Wang and P. Mukherjee, 2004 Astroph. J. {\bf 606} 654 \\
                       S. Nesseris and L. Perivolaroupolos, 2004 Phys. Rev. D {\bf 70} 043531\\
                       T. Roy Choudhury and T. Padmanabhan, 2005 Astron. Astrophys. {\bf 429} 807\\
                       Y. Gong, 2005 Int. J. Mod. Phys. D {\bf 14} 599  \\
                       C. Wetterich, 2004 Phys. Lett. B {\bf 594} 17 \\
                       P. Wu and H. Yu, 2006 Phys. Lett. B {\bf 643}
                       315\\
                       Z. K. Guo, N. Ohta and Y. Z. Zhang, 2005 Phys. Rev. D {\bf 72} 023504\\
                       Z. K. Guo, N. Ohta and Y. Z. Zhang, 2007 Mod. Phys. Lett. A {\bf 22} 883 \\
                       J. Simon, L. Verde and R. Jimenez, 2005 Phys. Rev. D {\bf 71} 123001
\bibitem{Wang2001} Y. Wang and G. Lovelace, 2001 Astroph. J. {\bf 562} L115 \\
                   T. D. Saini, 2003 Mon. Not. Roy. Ast. Soc. {\bf 344} 129\\
                   A. Daly and S. G. Djorgovsky, 2003 Astroph. J. {\bf 597} 9\\
                   R. A. Daly and S. G. Djorgovsky, 2004 Astroph. J. {\bf 612} 652\\
                   R. A. Daly and S. G. Djorgovsky, 2006 astro-ph/0609791\\
                   Y. Wang and M. Tegmark, 2004 Phys. Rev. Lett. {\bf 92} 241302 \\
                   Y. Wang and M. Tegmark, 2005 Phys. Rev. D {\bf 71} 103513 \\
                   R. A. Daly and S. G. Djorgovsky, 2005 astro-ph/0512576 \\
                   D. Huterer and A. Cooray, 2005, Phys. Rev. D {\bf 71} 023506 \\
                   D. Huterer and G. Starkman, 2003, Phys. Rev. Lett. {\bf 90} 031301\\
                   S. Fay and R. Tavakol, 2006 Phys. Rev. D {\bf 74} 083513 
\bibitem{Shaf2006} A. Shafieloo, U. Alam, V. Sahni and A. Starobinsky, 2006 Mon. Not. R. Astron. Soc. {\bf 366} 1081
\bibitem{Shaf2007} A. Shafieloo, 2007 Mon. Not. R. Astron. Soc. {\bf  380}  1573
\bibitem{Eisenstein2005}D. J. Eisenstein, et al.,  2005 Astrophys. J. {\bf 633} 560
\bibitem{Nesseris2006}S. Nesseris and L. Perivolaropoulos, 2007 J. Cosmol. Astropart. Phys. {\bf 01} 018 
\bibitem{Alam2006} U. Alam, V. Sahni and A. Starobinsky, 2007 J. Cosmol. Astropart. Phys. {\bf 02} 011 
\bibitem{WuYu2007} P. Wu and H. Yu, 2007 J. Cosmol. Astropart. Phys. {\bf 10} 014
\bibitem{Davis2007} T. M. Davis,  et al., arXiv:astro-ph/0701510\\
                 Barger V,  Gao Y and Marfatia D, 2007 Phys.Lett. B {\bf 648}  127
\bibitem{WMAP3}   D. N. Spergel, et al., 2007 Astrophys. J. Supp. {\bf 170}  377 
\bibitem{Gong} Y. G. Gong, 2005 Class. Quantum Grav. {\bf 22} 2121\\
               Y. G. Gong and Y.Z. Zhang, 2005 Phys. Rev. D {\bf 72} 043518 \\
               M. Chevallier and D. Polarski, 2001 Int. J. Mod. Phys. D {\bf 10} 213 \\
               Y.  Gong and A. Wang, 2007 Phys. Rev. D {\bf 75} 043520


\end{thebibliography}
\end{document}